\documentclass[%
 reprint,
nofootinbib,
 amsmath,amssymb,
 aps,
prd,
]{revtex4-1}

\usepackage{amssymb,amsmath}
\usepackage{euscript}
\usepackage{mathtools} 
\usepackage[usenames,dvipsnames]{color}
\usepackage{graphicx}
\usepackage{dcolumn}
\usepackage{bm}
\usepackage{hyperref}

\newcommand{\Fnu}{\tilde{F}_\nu}

\newcommand{\Bnu}{\tilde{B}_\nu}

\newcommand{\Gij}[2]{\prescript{#1}{#2}{\mathcal{G}}}

\newcommand{\betahat}{\hat{\bm \beta}}
\newcommand{\dhat}{\hat{\bm d}}
\newcommand{\gammahat}{\hat{\bm \gamma}}

\begin{document}

\preprint{APS/123-QED}

\title{Beyond the Boost:\\
	 Measuring the intrinsic dipole of the CMB using the spectral distortions of the monopole and quadrupole}

\author{Siavash Yasini}
\email{yasini@usc.edu} 
\author{Elena Pierpaoli}%
 \email{pierpaol@usc.edu}
\affiliation{%
 Physics \& Astronomy Department, University of Southern California, Los Angeles, California 91106\\
}%

\date{\today}

\begin{abstract}
 We present a general framework for accurate spectral modeling of the low multipoles of the cosmic microwave background (CMB) as observed in a boosted frame. In particular, we demonstrate how spectral measurements of the low multipoles can be used to separate the motion-induced dipole of the CMB from a possible intrinsic dipole component. In a moving frame, the leakage of an intrinsic dipole moment into the CMB monopole and quadrupole induces spectral distortions with distinct frequency functions that respectively peak at 337 GHz and 276 GHz. The leakage into the quadrupole moment also induces a geometrical distortion to the spatial morphology of this mode. The combination of these effects can be used to lift the degeneracy between the motion-induced dipole and any intrinsic dipole that the CMB might possess. Assuming the current peculiar velocity measurements, the leakage of an intrinsic dipole with an amplitude of $\Delta T = 30\mu$K into the monopole and quadrupole moments will be detectable by a \emph{PIXIE}--like experiment at $\sim 40~$nK ($2.5\sigma$) and $\sim 130~$nK ($11\sigma$) level at their respective peak frequencies. 
 
\end{abstract}

\maketitle


\emph{Introduction}. The measurements of the COBE/FIRAS instrument show that the intensity of the Cosmic Microwave Background (CMB) has an almost perfect blackbody spectrum \cite{Fixsen:1996nj}. Even though in a frame moving with respect to the CMB the observed intensity is effectively a blackbody in every direction, the intensity harmonic multipoles in this frame generally contain frequency spectral distortions. These distortions are a result of the leakage of the nearby multipoles into each other due to the aberration and Doppler effects \cite{Yasini:2016pby,Challinor2002,Chluba:2016bvg,Chluba:2004cn,Dai2014}.
The most prominent motion-induced leakage component is that of the monopole into the dipole (i.e. kinematic dipole). The kinematic dipole has a frequency dependence identical to a differential blackbody spectrum which makes it degenerate with any intrinsic (or non-kinematic) dipole that the CMB might possess. Current modeling of the CMB dipole only includes the leakage of the monopole, but ignores any intrinsic dipole component as well as other kinematic corrections to this mode (e.g. the leakage of the quadrupole).  
Here we present an accurate description of the frequency spectrum of the low multipoles of CMB and show how the kinematic (motion-induced) corrections to these modes can be used by the next generation of CMB surveys to lift the dipole degeneracy.

A kinematic dipole is not the only observational consequence of our motion with respect to the CMB.  The motion-induced leakage of the  intensity multipoles into each other causes a \emph{boost coupling} between the nearby multipoles.  Measuring this boost coupling in a wide range of harmonic modes can actually lead to an independent measure of the peculiar velocity of an observer with respect to the CMB \cite{Aghanim:2013suk,Notari:2011sb,Amendola2010,Kosowsky2010}.  In the CMB rest frame, all motion-induced effects (including the kinematic dipole and the boost coupling) vanish; however, there is no compelling reason for us to believe that the intrinsic dipole moment of the CMB in this frame is precisely zero.  

It has been shown that in a flat $\Lambda$CDM universe with adiabatic initial perturbations, the intrinsic dipole of the CMB is strongly suppressed \cite{Erickcek:2008jp,Zibin:2008fe}. For this reason, the intrinsic dipole of the CMB is usually either ignored or set to zero, and the observed dipole of the CMB is interpreted entirely as a  kinematic effect. This results in a peculiar velocity of $\beta\equiv v/c= 0.00123$ in the direction $\betahat=(264^\circ,48^\circ)$ in galactic coordinates \cite{Kogut:1993ag}. If the observed dipole moment only has a kinematic origin, it can be used to define a natural rest frame for CMB (namely, the frame in which the whole dipole vanishes). However, unintended subtraction of an existent non-kinematic dipole in this process will result in obtaining an incorrect CMB rest frame. This can in turn lead to unexpected anomalies, such as the observed power and parity asymmetries in the CMB \cite{Naselsky:2011jp,Freeman:2005nx} and the mismatch between the CMB rest frame and the matter rest frame \cite{Ma:2010ps,Gibelyou:2012ri,Turner:1991dn,Kashlinsky:2008ut}. Studying the angular variance of the Hubble parameter over different redshifts (in the CMB dipole-inferred frame) also indicates the presence of a non-kinematic dipole component in the CMB \cite{Wiltshire:2012uh,McKay:2015nea}. 
Furthermore, since isocurvature initial perturbations, and multi-field inflationary scenarios typically invoke a non-negligible intrinsic dipole moment, a detection of this component could have important implications for pre-recombination physics \cite{Mazumdar:2013yta,Lyth:2013vha,Roldan:2016ayx,Dai:2013kfa,Mathews:2014fwa}.

Recently the Planck team has obtained an independent value for the peculiar velocity of the solar system using the boost coupling of the CMB multipoles. Their result $\beta=0.00128 \pm0.00026$(stat.)$\pm 0.00038$(syst.)  \cite{Aghanim:2013suk} is consistent with the kinematic interpretation of the dipole and shows that most of the dipole that we observe is induced by our peculiar motion. However, the error bars still allow for a non-kinematic dipole component that remains to be measured.

In this letter we show how the kinematic and non-kinematic dipoles can be separated by measuring the motion-induced spectral distortions in the observed low multipoles of the CMB in our local frame. 
Future microwave surveys, such as \emph{PIXIE} with a sensitivity of 5~Jy/sr, will be able to measure these effects with high precision.

\emph{Lorentz boosting the CMB}. We define the rest frame of the CMB as the frame in which its kinematic dipole (the leakage of the monopole into the dipole) vanishes.\footnote{Indeed, in this frame all the other kinematic effects including the boost coupling and the ones that we are about to discuss will vanish as well.} We still allow the CMB to have a non-kinematic dipole in this frame. Then we argue that the full frequency spectrum of the low intensity multipoles in the boosted frame can be exploited to separate the intrinsic dipole from the kinematic part induced by a boost. 
We assume that the CMB frequency spectrum in its rest frame can be described as a pure blackbody by neglecting any pre-recombination and secondary $\mu$- and $y$-distortions (see Fig. 12 in \cite{Kogut:2011xw}, also \cite{Balashev:2015lla}).  
In this frame, we expand the intensity and the thermodynamic temperature in spherical harmonic multipoles as 

\begin{equation}\label{intensity_expansion}
I_{\nu_{cmb}}(\gammahat_{cmb}) = \sum_{\ell=0}^{\infty}\sum\nolimits_{m}^{\ell} a_{\ell m}^{I_{cmb}}(\nu_{cmb})~ Y_{\ell m}(\gammahat_{cmb})
\end{equation}
and 
\begin{equation}\label{T_expansion}
T(\gammahat_{cmb}) = \sum_{\ell=0}^{\infty}\sum\nolimits_{m}^{\ell} a_{\ell m}^{T_{cmb}}~ Y_{\ell m}(\gammahat_{cmb}),
\end{equation}
where the sum notation $\sum_{m}^{\ell}$ is shorthand for $\sum\limits_{m=-\ell}^{\ell}$. The frequency dependence of the intensity harmonic coefficients for a blackbody---with an average temperature $T_0$---can be expanded to first order in thermodynamic temperature harmonics as 

\begin{subequations}\label{alm(I)_to_alm(T)}
	\begin{align}
	a^{I_{cmb}}_{00}(\nu)&=\tilde{B}_\nu(T_0)~a^{T_{cmb}}_{00},\\
	a^{I_{cmb}}_{\ell m}(\nu)&=\tilde{F}_\nu(T_0)~a^{T_{cmb}}_{\ell m}\quad(\ell>0),
	\end{align}
\end{subequations}
where $\Bnu(T_0)\equiv T_0^{-1} B_\nu(T_0)$, $B_\nu(T)\equiv\frac{2h\nu^3}{c^2}\frac{1}{e^{h\nu/kT}-1}$ is the blackbody spectrum and $\tilde{F}_\nu(T_0)\equiv \Bnu(T_0)f(x)$ is the differential blackbody spectrum with $f(x)\equiv~\frac{xe^{x}}{e^{x}-1}$ and $x=h\nu/kT_0$.

 In order to find the observed multipoles in the boosted frame we use the Lorentz invariance of $I_\nu/\nu^3$ to write the observed incoming intensity along the line-of-sight unit vector $\gammahat$ at frequency $\nu$ as

\begin{equation}\label{I_prime}
I_{\nu}( \gammahat)=\Big(\frac{\nu}{\nu_{cmb}}\Big)^3 I_{\nu_{cmb}}( \gammahat_{cmb}),
\end{equation}
where 
\begin{equation}\label{doppler_freq}
\nu_{cmb}=\Big(\frac{1-\beta \mu}{\sqrt{1-\beta^2}}\Big)\nu
\end{equation}
 and
  \begin{equation}\label{aberration}
 \gammahat_{cmb}=\Big(\frac{(1-\sqrt{1-\beta^2})\mu-\beta}{1-\beta \mu}\Big )\betahat+\Big (\frac{\sqrt{1-\beta^2}}{1-\beta \mu}\Big)\gammahat
 \end{equation}
 are the frequency and line-of-sight unit vector in the CMB rest frame and $\mu=\gammahat \cdot \betahat$.  Equations \eqref{doppler_freq}  and \eqref{aberration} respectively represent the Doppler  and aberration effects. Expanding both sides of Eq. \eqref{I_prime} in harmonic space allows us to find the observed multipoles in the moving frame as

\begin{multline}\label{alm_final}
a^{I}_{\ell' m'}(\nu) =\\
\sum_{\ell=0}^{\infty}\sum\nolimits_{m}^{\ell} \int \Big(\frac{\nu}{\nu_{cmb}}\Big)^3 a^{I_{cmb}}_{\ell m}(\nu_{cmb}) Y_{\ell m}(\gammahat_{cmb}) Y^*_{\ell' m'}(\gammahat) \text{d}^2 \gammahat.
\end{multline}
 Substituting Eqs. \eqref{doppler_freq} and \eqref{aberration} into \eqref{alm_final} will respectively result in the \emph{Doppler and aberration leakage} of the nearby multipoles into each other. To $n$-th order in $\beta$, the observed multipoles $a^{I}_{\ell' m'}(\nu) $ will have a contribution from $a^{I_{cmb}}_{\ell'\pm n, m'}(\nu) $ of the rest frame. This integral has been computed analytically in Ref.\cite{Yasini:2016pby}. We do not repeat the calculations here and only use the results hereafter. We also acquire the same notation for the frequency functions.

\emph{The boosted dipole}. First, we calculate the observed dipole in the moving frame to illustrate the dipole degeneracy problem. By setting $\ell'=1$ in Eq.~\eqref{alm_final} we find  (Eq. B.37 in Ref. \cite{Yasini:2016pby})

\begin{equation}\label{final_dipole}
	\begin{split}
		&a^{I}_{1m'}(\nu)=\overbrace{\vphantom{\int^1} \Fnu(T_0)a^{T_{cmb}}_{1m'}}^\text{Intrinsic dipole}
		+\overbrace{\beta ~
			\frac{2\sqrt{\pi}}{3}Y^*_{1m'}(\hat{\bm{\beta}})\Fnu(T_0)a^{T_{cmb}}_{00}}^\text{Kinematic dipole}\vspace{0.5em}\\
		&+\beta \sum_{m,n}^{2,1}
		\Gij{1}{0}^{2m}_{1m'}(\betahat)\Fnu^{(11)}(T_0)a^{T_{cmb}}_{2m}\vspace{0.5em}\\
		&+\beta ~\sum\limits_{m,n}^{2,1} \Gij{0}{1}^{2m}_{1m'}(\betahat)
		\Fnu(T_0)a^{T_{cmb}}_{2 m}\\
		&+O(\beta^2),
	\end{split}
\end{equation}
where $\Fnu^{(11)}(T)= \Fnu(T)(g(x)-1)$ with $g(x)\equiv~x \coth(x/2)$, while $	\Gij{1}{0}^{2m}_{1m'}(\betahat)$ and	$\Gij{0}{1}^{2m}_{1m'}(\betahat)$ are numerical factors of order $\sim1$. 

The first term in Eq. \eqref{final_dipole} is the intrinsic dipole of the CMB with the differential blackbody spectrum $\Fnu(T_0)$. The second term is what is normally identified as the kinematic dipole, which is a result of the Doppler leakage of the monopole into the observed dipole moment. Notice that the frequency dependence of this terms is identical to the intrinsic dipole which makes the two components degenerate. 
The third and the fourth terms are respectively the Doppler and aberration leakages of the quadrupole into the dipole. 
These terms have never been considered in the analysis of the CMB dipole. 

\begin{figure}
\centering
\includegraphics[width=1.0\linewidth]{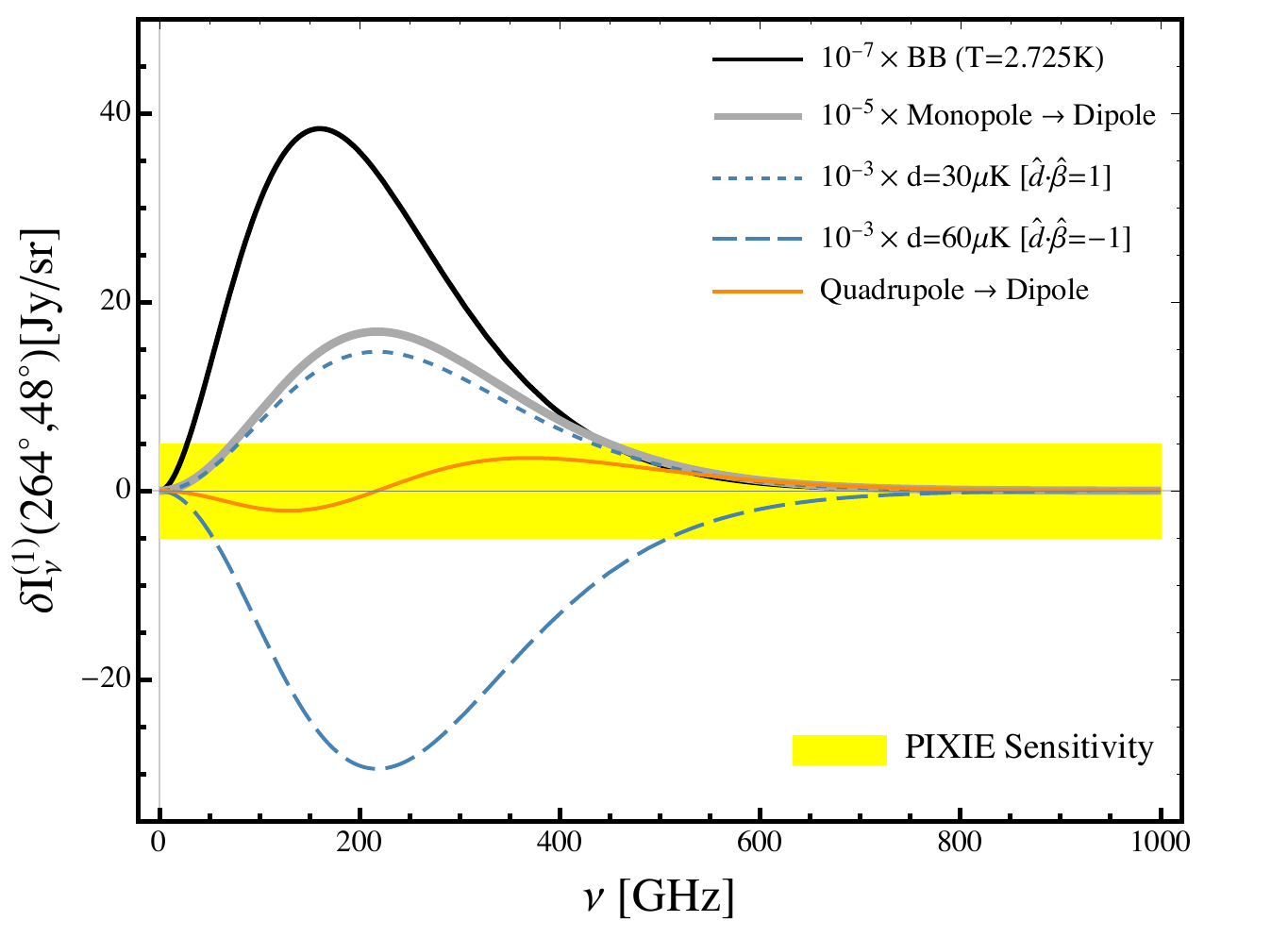}
\caption{The CMB dipole constituents observed in a moving frame with $\beta=0.00128$ and $\betahat=(264^\circ,48^\circ)$. 
	The intrinsic dipoles d30 
	and d60 
	have identical frequency functions as the kinematic dipole. 
	The average T=2.725 blackbody spectrum (\emph{solid black}) 
	is depicted in all plots for reference. }
\label{fig:dipolecorrections}
\end{figure}

In order to build some intuition, instead of working with the $a^{T_{cmb}}_{1,m}$ coefficients, we parametrize the three degrees of freedom for the intrinsic dipole in terms of an amplitude and two angles via the definition 

\begin{equation}\label{dipole_coeffs}
a^{T_{cmb}}_{1,m}\equiv \frac{4\pi}{3}dY^*_{1,m}(\theta_d,\phi_d).
\end{equation}
We define the dipole vector $\vec{\bm d}=d \dhat$ where $d$ and  $\dhat\equiv(\theta_d,\phi_d)$ are the amplitude and direction of the maximum of the dipole on the sky. 

With this new definition, we set out to study the observable effects of an intrinsic dipole of order $\sim10^{-5}$ on the local dipole, monopole and quadrupole of the CMB. In order to gauge the expected magnitude of the effect we will consider two different dipoles with the amplitudes
$d=30\mu$K and $d=60\mu$K (motivated by Ref. \cite{Erickcek:2008jp} Eqs. 31-33). We will refer to these dipoles respectively as d30 and d60. 

The observed dipole intensity in the direction $(\theta,\phi)$ is defined as $\delta I^{(1)}_\nu (\theta,\phi) \equiv \sum_{m'}^{1} a^I_{1 m'}(\nu) Y_{1 m'}(\theta,\phi)$. Fig. \ref{fig:dipolecorrections} shows the contribution of each term in Eq.~\eqref{final_dipole} to $\delta I^{(1)}_\nu (\theta_\beta,\phi_\beta)$ at different frequencies.
Unless the intrinsic dipole is much larger than the one we chose, the dominant term in this equation is the leakage of the monopole into the dipole (kinematic dipole) with the thermodynamic temperature $\delta T^{(1)} \equiv \delta I^{(1)}_\nu/ \Fnu(T_0) = 3.35$ mK. The next order contribution is due to the intrinsic dipole with the same frequency function as that of the kinematic dipole. 
The leakage of the quadrupole into the dipole is a motion-induced effect which does not depend on the intrinsic dipole at all. Since this term has a different frequency dependence, technically it could be used as an independent measure of $\beta$. However, the peak amplitude of this component---assuming the observed value of the quadrupole as input---is lower than the sensitivity of \emph{PIXIE}, and therefore it is not likely to be useful for lifting the dipole degeneracy. Nevertheless, this extra leakage component should be taken into account for a precise analysis of the observed dipole in the future CMB surveys. 

Now we show how the dipole degeneracy can be removed by looking at the motion induced spectral distortions in the dipole's neighbors: the monopole ($\ell'=0$) and the quadrupole ($\ell'=2$).

\emph{The boosted monopole}. 
Using Eq. \eqref{alm_final}, it is easy to find the monopole of the CMB in a boosted frame  (Eq. B.36 in Ref. \cite{Yasini:2016pby})
	\begin{equation}\label{final_monopole}
	\begin{split}
	a^{I}_{00}(\nu)=& \Bnu(T_0)a^{T_{cmb}}_{00}+\beta^2\Bnu^{(20)}(T_0)a^{T_{cmb}}_{00}\\
	+\beta&~\sum\nolimits_{m}^{1}
	\frac{2\sqrt{\pi}}{3}Y_{1m}(\hat{\bm{\beta}})\Fnu^{(11)}(T_0)a^{T_{cmb}}_{1 m}\\
	-\beta&~ \sum\nolimits_{m}^{1} \frac{4\sqrt{ \pi}}{3}Y_{1m}(\betahat)\Fnu(T_0)a^{T_{cmb}}_{1 m}
	+O(\beta^2),
	\end{split}
	\end{equation}
with $\Bnu^{(20)}(T_0)=\frac{1}{6}\Fnu(T)(g(x)-3)$. Here the first term is the well known $T=2.725$ blackbody spectrum, the second term is the second order Doppler correction to the monopole, and the third and fourth terms are respectively the Doppler and aberration leakages of the dipole into the monopole.

\begin{figure}[t]
	\centering
	\includegraphics[width=1.0\linewidth]{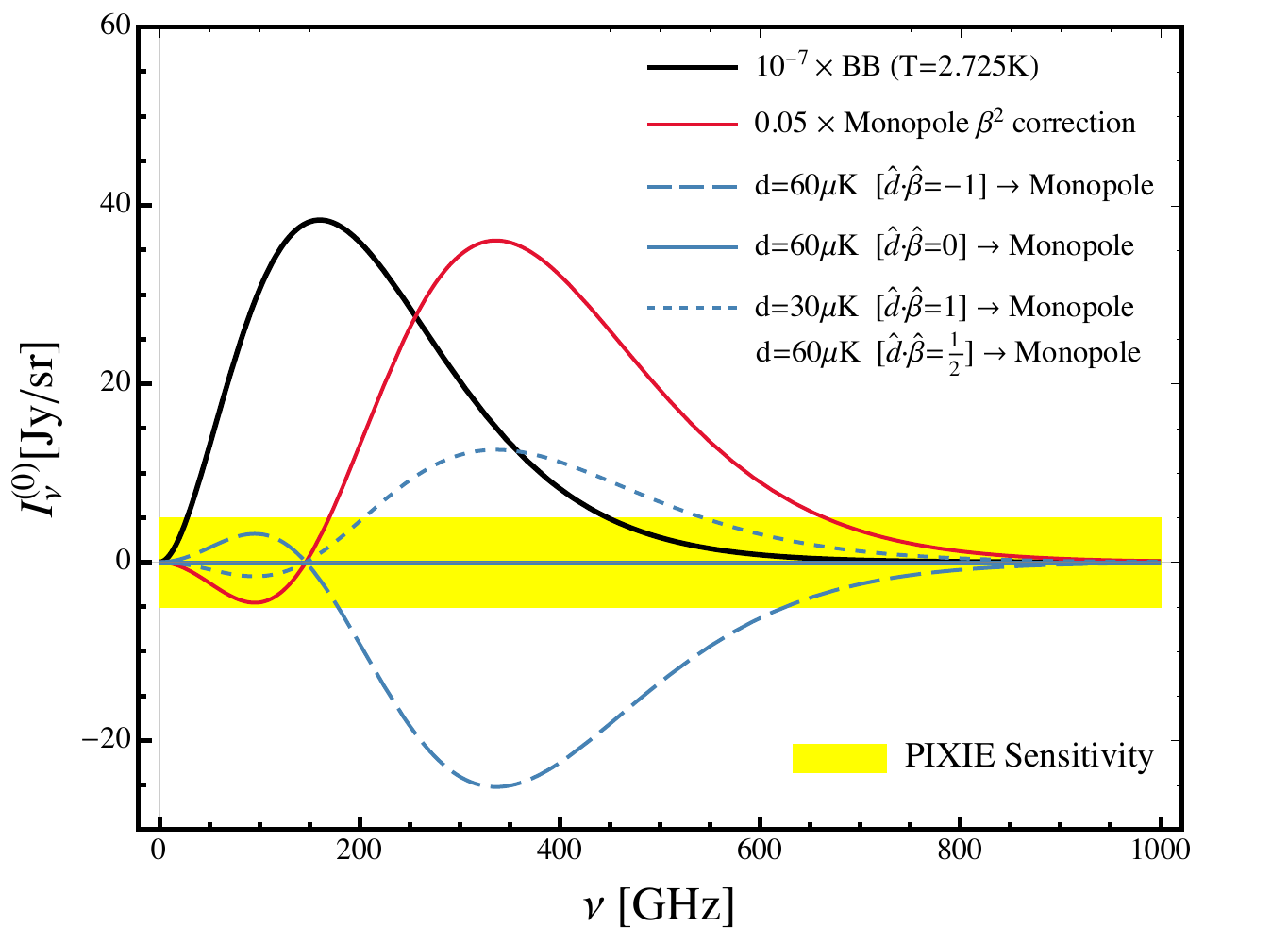}
	\caption{The motion induced spectral distortions of the observed CMB monopole. 
	 }
	\label{fig:monopolecorrections}
\end{figure}

The observed monopole intensity $I^{(0)}_\nu (\theta,\phi)= a^{I}_{00}(\nu) Y_{00}(\theta,\phi)=a^{I}_{00}(\nu)/2\sqrt{\pi}$ is plotted in Fig. \ref{fig:monopolecorrections} for different amplitudes and orientations of the intrinsic dipole. 
Using Eq. \eqref{dipole_coeffs}, we can rewrite Eq. \eqref{final_monopole} as 

\begin{equation}\label{monopole_simplified}
\delta I_\nu^{(0)}= \Bnu(T_0) T_0 + \beta \Bnu^{(20)}(T_0) [\beta T_0 + 2 d (\dhat \cdot \betahat)].
\end{equation}
Since the frequency dependence of the intrinsic monopole $T_0$ is different from the motion induced terms, it can be fit and measured separately. Since the motion-induced spectral distortions depend the combination of the kinematic dipole ($\beta T_0$) and the projection of the intrinsic dipole along the direction of motion ($d (\dhat \cdot \betahat$)), it might seem like these two components still remain degenerate. However, combining this with the observed dipole in $\betahat$ direction (with the quadrupole leakage term dropped, assuming it's negligible)
 \begin{equation}\label{dipole_simplified}
\delta I_\nu^{(1)}(\betahat)= \Fnu(T_0)[ \beta T_0 + d (\dhat \cdot \betahat)],
\end{equation}
reveals that the monopole spectral distortion adds an independent equation that allows one to separate $\beta T_0$ and $d (\dhat \cdot \betahat)$.  



Since the leakage of the dipole into the monopole only depends on the projection of $\vec{\bm d}$ along $\vec{\bm \beta}$ (see Fig. \ref{fig:monopolecorrections}), only by looking at the monopole alone one cannot find all three components of $\vec{\bm d}$; there remains an azimuthal degeneracy between the two vectors. Now we show that the leakage of the intrinsic dipole into the quadrupole can be exploited to find both the amplitude and direction of $\vec{\bm d}$.  

\begin{figure}[!t]
	\centering
	\includegraphics[width=1.0\linewidth]{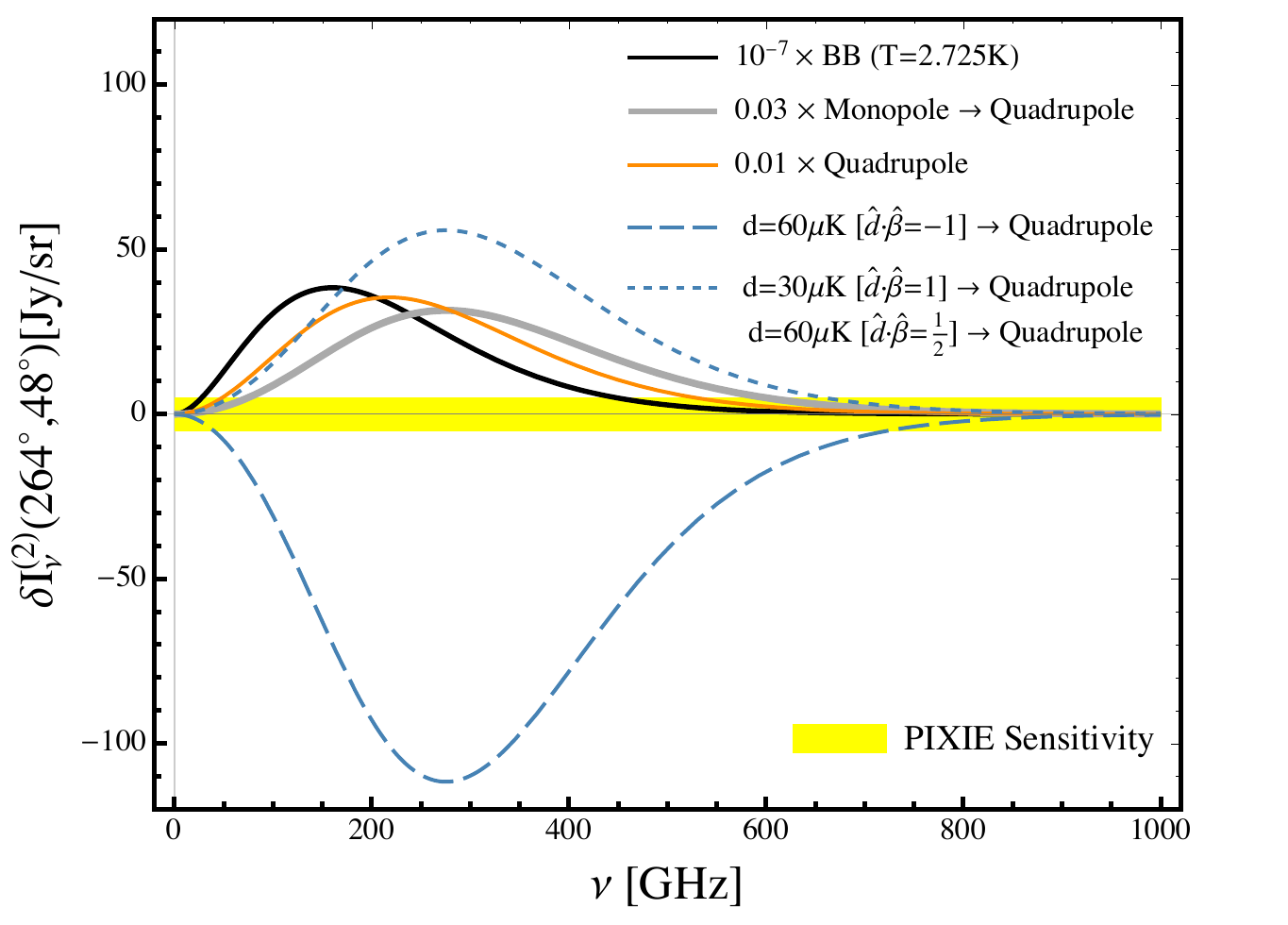}
	\caption{The motion induced spectral distortions of the observed CMB quadrupole. Since the leakage of the octupole has a different frequency function compared to the other components, we have assumed that this term can be identified and subtracted and therefore is not shown here. In this specific direction in the sky, the leakage of the d30  is not distinguishable from a d60 with the same projection along $\betahat$ (\emph{short dashed blue}). However, the amplitude of these two leakage components are different at other lines of sight (see figure \ref{fig:dipole_to_quadrupole}).
	}
	\label{fig:quadrupolecorrections}
\end{figure}

\emph{The boosted quadrupole}. 
In a boosted frame the intrinsic dipole also leaks into the observed quadrupole   (Eq. B.38 in Ref. \cite{Yasini:2016pby})
\begin{equation}\label{final_quadrupole}
\begin{split}
a^{I}_{2 m'}(\nu)=& \Fnu(T_0)a^{T_{cmb}}_{2 m'}+\\
+&\beta^2 ~\frac{2\sqrt{\pi}}{5}Y^*_{2m'}(\betahat)\Bnu^{(22)}(T_0)a^{T_{cmb}}_{00}\\
+&\beta \sum\nolimits_{m,n}^{1,1}
\Gij{1}{0}^{1m}_{2m'}(\betahat)\Fnu^{(11)}(T_0)a^{T_{cmb}}_{1 m}\\
+&\beta \sum\nolimits_{m,n}^{1,1} \Gij{0}{1}^{1m}_{2m'}(\betahat)
\Fnu(T_0)a^{T_{cmb}}_{1 m}\\
+&\beta \sum\nolimits_{m,n}^{3,1} \Gij{1}{0}^{3m}_{2m'}(\betahat)\Fnu^{(11)}(T_0)a^{T_{cmb}}_{3 m}\\
+&\beta \sum\nolimits_{m,n}^{3,2} \Gij{0}{1}^{3m}_{2m'}(\betahat)
\Fnu(T_0)a^{T_{cmb}}_{3 m}+O(\beta^2),
\end{split}
\end{equation}
where $\Bnu^{(22)}\equiv\frac{1}{3}\Fnu(T)g(x)$. The largest term here is the intrinsic quadrupole, followed by the leakage of the monopole into the quadrupole (the second term).  The third and fourth terms represent the Doppler and aberration leakage of the intrinsic dipole into the quadrupole\footnote{The frequency function of the dipole leakage is different from Ref. \cite{Kamionkowski:2002nd} which does not account for the aberration effect.}. 
Fig. \ref{fig:quadrupolecorrections} shows the contribution of different terms in Eq. \eqref{final_quadrupole} to the observed quadrupole intensity $\delta I^{(2)}_\nu (\theta,\phi) \equiv \sum_{m'}^{2} a^I_{2 m'}(\nu) Y_{2 m'}(\theta,\phi)$. 

\begin{figure}[!t]
	\centering
	275 GHz
	\includegraphics[width=0.9\linewidth]{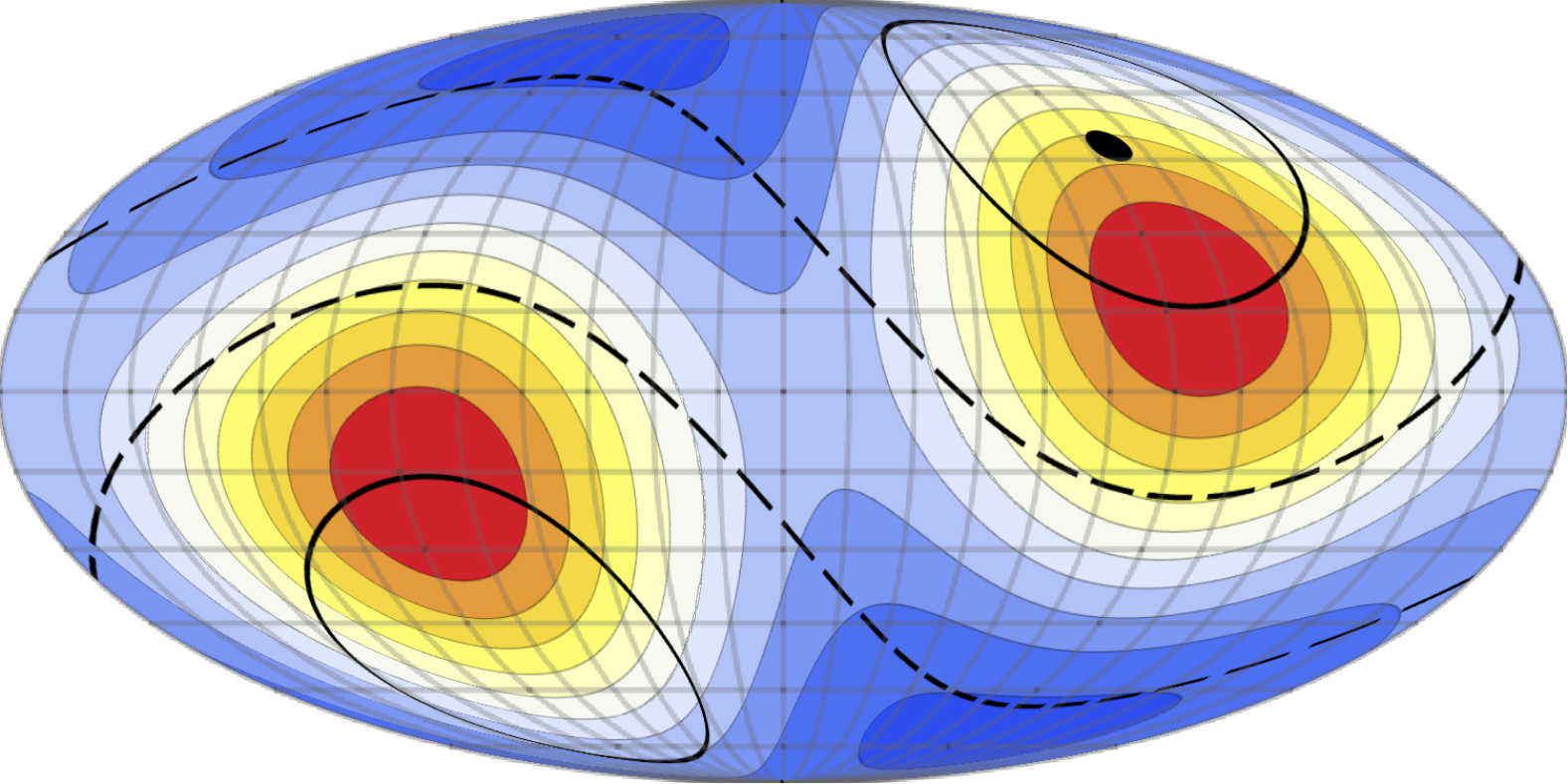}
	\includegraphics[width=1.0\linewidth]{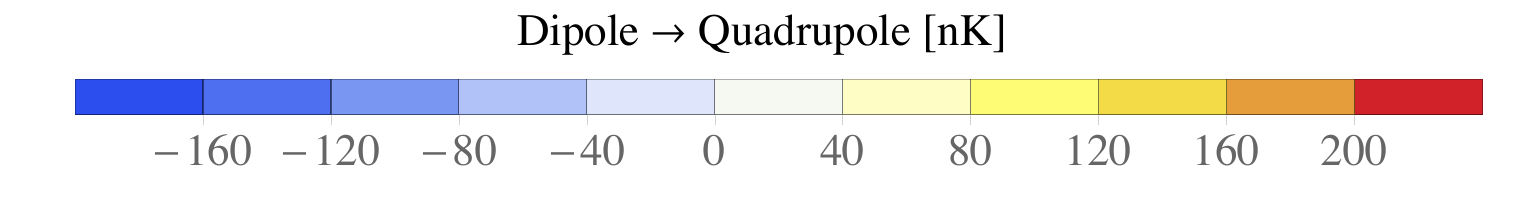}
	\caption{Mollweide projection of the leakage of an intrinsic dipole d60 with $\dhat \cdot\betahat=1/2$ into the observed quadrupole. The \emph{solid (dashed) black} lines are the -40nK (80nK) contour lines for the leakage of a smaller intrinsic dipole d30 with a different orientation $\dhat=\betahat$. Even though the two dipole leakage components have the same amplitude along the $\betahat$ direction (\emph{black dot}), their spatial morphology is different over the whole sky.  }
	\label{fig:dipole_to_quadrupole}
\end{figure}

Note that the leakage of the intrinsic dipole d30 aligned with $\betahat$ induces the same signal in the line of sight direction $\betahat$, as the d60 dipole with $\dhat \cdot \betahat=1/2$. However, in contrast to the case of the monopole, the spatial morphology of the dipole leakage into the quadrupole is not uniform over the whole sky and depends on $\dhat$. Fig. \ref{fig:dipole_to_quadrupole} shows this difference for two cases of dipoles with the same parallel component along $\betahat$ but different $\dhat$s. Therefore, the whole sky map of the leakage component can be used to lift the degeneracy between the amplitude and the orientation of the dipoles. The dipole leakage into the quadrupole adds five independent equations to Eq. \eqref{final_dipole} and \eqref{final_monopole} which, combined together,\label{key} are more than enough for simultaneous determination of $\vec{\bm d}$ and $\vec{\bm \beta}$. 

\emph{Discussion}. Future generation of microwave experiments are going to make accurate measurements of the frequency spectrum of the CMB. We presented a framework for accurate spectral modeling of the low multipoles of the CMB in a moving frame that should be considered in the future CMB surveys. In particular, we showed how measuring the spectral distortions in the CMB multipoles can be used to distinguish between the motion-induced and intrinsic dipole components of the CMB.  The main idea is that our peculiar motion with respect to the CMB rest frame causes the low multipoles of the CMB to leak into each other. These leakage components induce distinct frequency distortions that can be used to determine both the amplitude and orientation of a possible intrinsic dipole in the CMB and separate it from the kinematic dipole.

 
The main challenge in the detection of these signals is likely imposed by  foregrounds which however, will be accurately mapped by the next generation of CMB surveys with a multitude of frequency channels (e.g. \emph{PIXIE} with 400 channels). 
\emph{PIXIE} will be able to detect the leakage of a $30\mu$K  dipole into the monopole and quadrupole at the peak frequencies (337 GHz and 276 GHz) with $\sim 2.5\sigma$ and $11 \sigma$.

\emph{Acknowledgments.} We sincerely thank Jens Chluba, Alessio Notari, Miguel Quartin and Adrianna Erickcek for incredibly helpful discussions. We also thank the anonymous referee for their invaluable comments on our manuscript. 

\bibliographystyle{apsrev4-1}	
\bibliography{kSZPol}

\appendix
\numberwithin{equation}{section}	

	

	\section{Definition of the reference frame}
	
	It is commonly assumed that the CMB rest frame is the one in which the whole \emph{local} dipole moment (as observed in the CMB temperature) vanishes. However, it is plausible that primordial physical processes caused the imprint of a \emph{global} dipole signature on the CMB. 
	The frame in which all the dipole is of the latter nature is defined as the CMB rest frame. 
	
	Let us assume that in this frame the intensity in every direction is represented by a prefect black-body:
	
	\begin{equation}
	I_{\nu_\text{cmb}}(\gammahat_\text{cmb})=\frac{2 h }{c^2}\frac{\nu_\text{cmb}^3}{e^{h\nu_\text{cmb}/k T^\text{cmb}(\gammahat_\text{cmb})}-1}.
	\end{equation}
	Let us also assume that the  $a_{\ell m}^{T_\text{cmb}}$ multiples are  all uncorrelated and  Gaussian distributed.
	
	Let us now consider an observer in motion with respect to the CMB frame.
	The effects of a boost on the observed intensity 
	\begin{equation}\label{I_prime}
	I_{\nu}( \gammahat)=\Big(\frac{\nu}{\nu_{cmb}}\Big)^3 I_{\nu_{cmb}}( \gammahat_{cmb}),
	\end{equation}
	can be described via the adoption of an effective temperature
	\begin{equation}
	I_\nu(\gammahat)=\frac{2 h }{c^2}\frac{\nu^3}{e^{h\nu/k T^\text{eff}(\gammahat)}-1},
	\label{eq:InuT}
	\end{equation}
	with
	\begin{equation}\label{temperature_transformation}
	T^\text{eff}(\gammahat)=\frac{\sqrt{1-\beta^2}}{1-\beta \betahat \cdot \gammahat}T^\text{cmb}(\gammahat_\text{cmb}).
	\end{equation}
	where $T^\text{cmb}$ is the temperature of the CMB in its rest frame. 
	So the 
	effects of the motion  can be reinterpreted as a change in temperature in every direction. Therefore by looking at the CMB in a single direction one cannot distinguish between an intrinsic temperature fluctuation and a motion-induced one. However, the difference between the two temperatures in equation \eqref{temperature_transformation} is that the harmonic multipoles of $T^\text{eff}$ are correlated to each other due to the boost, but the multipoles of $T^\text{cmb}$ are not. By integrating Eq. \eqref{temperature_transformation} over $Y^*_{\ell' m'}(\gammahat)$ one can transform this equation in harmonic space \cite{Challinor2002,Chluba2011}
	
	\begin{equation}\label{Temperature_Kernel}
	a^{T_\text{eff}}_{\ell' m'} = \sum_{\ell,m} \mathcal{K}^{\ell,m}_{\ell' m'}(\vec{\bm \beta}) a^{T_\text{cmb}}_{\ell m}.
	\end{equation}

	It has been shown \cite{Challinor2002,Yoho:2012am,Chluba2011,Kosowsky2010} that an observer in motion with respect to the CMB frame will observe a correlation in the $a_{\ell m}^{T_\text{eff}}$ multipoles.
	These correlations have been exploited by the Planck team in order to determine our local peculiar velocity $\beta=v/c$ with respect to the CMB \cite{Aghanim:2013suk}.
	
	However, this is not the only observational feature of the motion. In Ref. \cite{Yasini:2016pby} we show that, due to the aberration and Doppler effects---which link the direction of motion with specific frequency and direction changes of the observed photons---the observed intensity multiples $a_{\ell m}^I$ in a moving frame will have a different frequency dependence and amplitude than the ones observed in the CMB rest frame.
	So effectively, despite the fact that the intensity in every direction is still effectively a black-body as indicated by Eq.\ref{eq:InuT}, observing the intensity multiples at various frequencies offers an alternative way  to infer our  peculiar motion with respect to the CMB rest frame, while also measuring the intrinsic intrinsic dipole. 
	The present letter focuses on the use of spectral distortions to disentangle the primordial CMB dipole from the one induced by our motion, by leveraging on the measurements of low--$\ell$ intensity multiples at various frequencies.
	
	The following explanatory note is aimed at easing the reader into the non-intuitive fact that a map which looks like a black-body in every direction (even when accounting for the observer's peculiar motion with respect to the CMB rest-frame)  may in fact present different frequency dependences  in the intensity multiples according to whether the observer is or is not in the CMB reference frame.

	\section{Intensity multipoles vs. temperature multipoles: non-linear effects}
	
	Let's first consider  how the  intensity multipoles connect to the temperature multiples for a sky that can be described in any direction as a black-body. These transformations would apply {\it no matter} which process produced that particular sky, so 
	they are valid both in the CMB  rest frame and in a different (moving) frame.

	By expanding the intensity to second order in temperature, one can easily find the nonlinear contribution of the temperature multipoles to the intensity monopole as
	
	\begin{align}\label{monopole_intensity}
	a^{I}_{00}(\nu) &=\tilde{B}_\nu(T_0) a^{T}_{00} + \tilde{G}_\nu (T_0)\sum_{\ell,m} \frac{1}{2 \sqrt{\pi}} | a^{T}_{\ell m}|^2,
	\end{align}
	where $\ell>0$, $\tilde{G}_\nu (T_0)=\frac{1}{2}T^{-1}_0(g(x)-1)\tilde{F}_\nu (T_0)$ and $T_0 = a^T_{00}/2\sqrt{\pi}$.
	
\begin{figure}[h]
		\centering
		\includegraphics[width=0.9\linewidth]{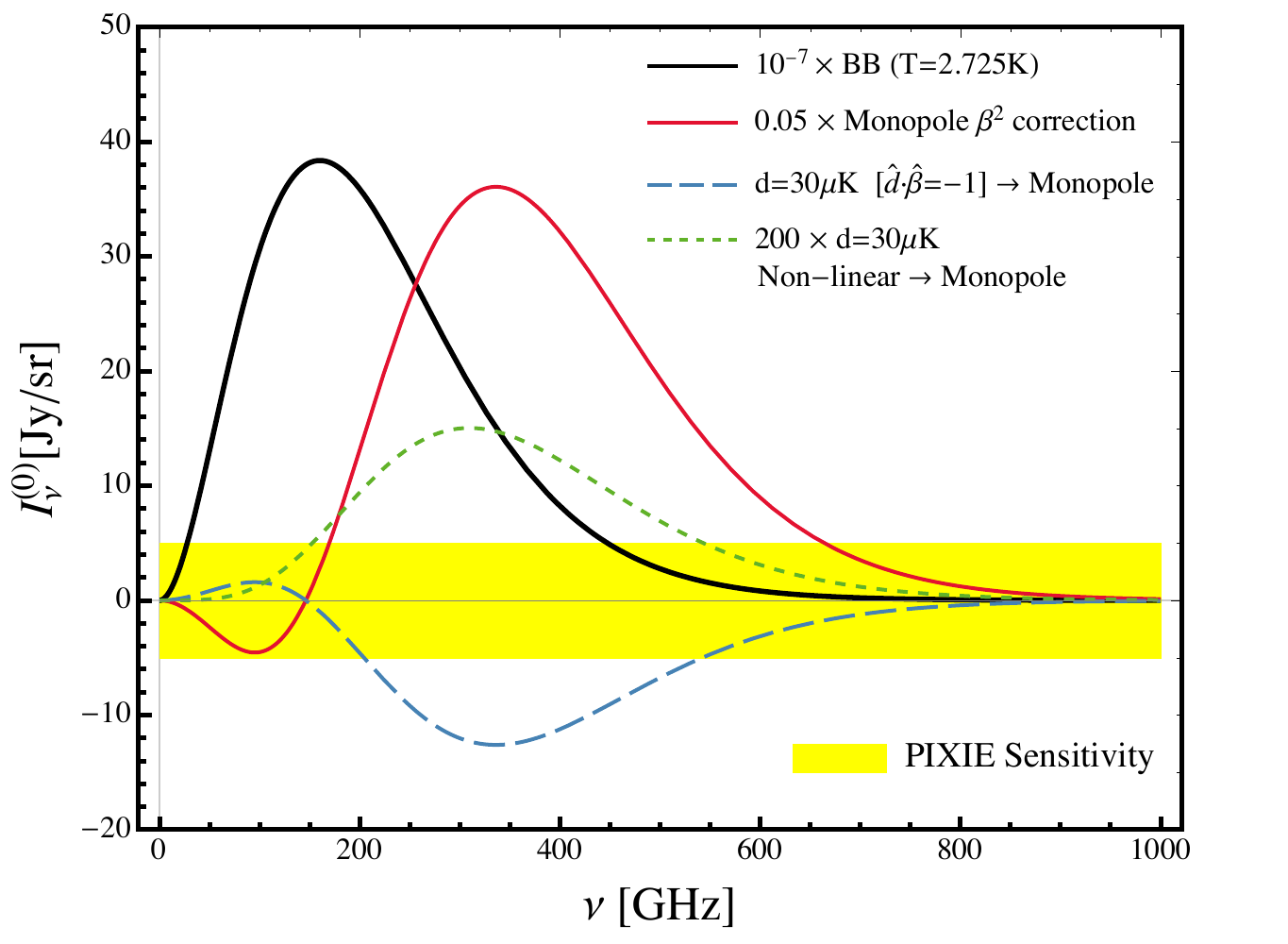}
		\caption{Nonlinear (second order in temperature expansion) contribution of a temperature dipole to the observed intensity monopole (see eq. C.7).}
		\label{fig:monopolecorrectionsd30uknonlinear}
	\end{figure}

	The frequency dependence for intensity multipoles in the CMB frame, to first order in $\Delta T$, are expected to be the ones in Eq. 3 of the letter. Corrections at higher orders in $\Delta T$ produce nonlinear effects which are typically about $10^{-5}$ times smaller. 
	Fig.\ref{fig:monopolecorrectionsd30uknonlinear} shows the nonlinear contribution of an intrinsic dipole in temperature to the intensity monopole.

	For completeness, we also provide the expression for nonlinear  (second order) temperature corrections to the higher multipoles of intensity ($\ell > 0$)
	
	\begin{align}\label{higher_multipole_intensity}
	a^{I}_{\ell m}(\nu) &=\tilde{F}_\nu(T_0) a^{T}_{\ell m} +\\\nonumber
	& \tilde{G}_\nu (T_0)\sum_{\ell',m'} \sum_{\ell'',m''} a^{T}_{\ell' m'}a^{T}_{\ell'' m''} \Delta_{\ell,m}(\ell',m';\ell'',m'') ,
	\end{align}
	where $\ell' , \ell'' >0$ and 
	\begin{multline}\label{Gaunt}
	\Delta_{\ell,m}(\ell',m';\ell'',m'') \equiv \int Y^*_{\ell m}(\gammahat) Y_{\ell' m'}(\gammahat)Y_{\ell'' m''}(\gammahat)  \text{d}^2 \gammahat \\
	=(-1)^{m}\sqrt{\frac{(2\ell+1)(2\ell'+1)(2\ell''+1)}{4\pi}} \\
	\times \begin{pmatrix}
	\ell& \ell'& \ell''\\
	0& 0& 0\\
	\end{pmatrix}
	\begin{pmatrix}
	\ell& \ell'& \ell''\\
	-m& m'& m''\\
	\end{pmatrix}.
	\end{multline}

	Leveraging on the basic and generally accepted knowledge presented in these first two sections, 
	in what follows we show the expected frequency dependences of the low-intensity multiples in a generic frame, in motion with respect to the CMB.
	
	\section{Understanding the monopole spectral distortions}
	\label{sec:refframes}

	As a pedagogical example, let's consider a CMB sky that only has an intrinsic temperature  monopole and dipole moments. In this scenario, an observer in the CMB frame  will only see a temperature monopole and dipole in their sky, but according to Eq. \eqref{Temperature_Kernel}, an observer in the moving frame will observe a monopole, dipole, quadrupole and so on. Let's only focus on the observed temperature monopole and dipole in the moving frame. From Eq. \eqref{Temperature_Kernel} we find 
	
	\begin{multline}\label{T_eff_monopole}
	a^{T_\text{eff}}_{00} = (1-\frac{\beta^2}{6}) a^{T_\text{cmb}}_{00} -\frac{2\sqrt{\pi}}{3} \beta \sum_{m} a^{T_\text{cmb}}_{1 m} Y_{1 m}(\betahat) \\+ O(\beta^2a^{T_\text{cmb}}_{1 m} ).
	\end{multline}
	Since the monopole does not get aberrated, the correction to the monopole term is only due to the Doppler effect, but the leakage of the dipole is due to both Doppler and aberration effects. Similarly, for the dipole we find
	\begin{multline}\label{T_eff_dipole}
	a^{T_\text{eff}}_{1 m} = \frac{2\sqrt{\pi}}{3} \beta  a^{T_\text{cmb}}_{00} Y^*_{1 m}(\betahat)  +  a^{T_\text{cmb}}_{1 m} \\ + O(\beta^2 a^{T_\text{cmb}}_{1 m}),
	\end{multline}
	where the leakage of the monopole is again only due to the Doppler effect. 
	
	For the sake of improving intuition, we use $a^{T_\text{eff}}_{00} =  2\sqrt{\pi} T^\text{eff}_0$, $a^{T_\text{cmb}}_{00} =  2\sqrt{\pi} T^\text{cmb}_0$, $a^{T_\text{cmb}}_{1m}=(4\pi/3)d Y^*_{1 m}(\dhat)$  and $\delta T^\text{eff}_{(1)}(\gammahat) =\sum_m a^{T_\text{eff}}_{1m} Y_{1 m}(\gammahat) $  to rewrite Eq. \eqref{T_eff_monopole} and \eqref{T_eff_dipole} as

	\begin{equation}\label{T_eff_monopole_simp}
	T^\text{eff}_0 = (1-\frac{\beta^2}{6}) T^\text{cmb}_0 -\frac{1}{3} \beta d (\dhat \cdot \betahat)
	\end{equation}
	\begin{equation}\label{T_eff_dipole_simp}
	\delta T^\text{eff}_{(1)}(\gammahat)  = ( \beta T^\text{cmb}_0 \betahat + d\dhat) \cdot \gammahat
	\end{equation}
	From Eq. \eqref{T_eff_dipole_simp} it would naively seem that any intrinsic dipole component $d$, could be absorbed into the kinematic dipole $\beta T^\text{cmb}_0 \betahat$. This may lead to the misconception that any primordial dipole the CMB may possess is indistinguishable from the motion-induce dipole. However, from Eq. \eqref{T_eff_monopole_simp}, it is obvious that a boost (leading to a kinematic dipole) changes the monopole temperature, but an  intrinsic dipole does not---since the temperature monopole and dipole are decoupled in the CMB rest frame.  In other words, in the CMB rest frame the intrinsic dipole does not have any effect on the monopole whatsoever. An intrinsic dipole changes the observed temperature monopole if and only if the observer is moving with respect to the CMB ($\beta \neq 0)$.

	Nevertheless, the monopole temperature is just one number, and any change either due to intrinsic or kinematic dipole in this number remains degenerate with the value of the monopole itself. However, as it is apparent from Fig.\ref{fig:monopolecorrectionsd30uknonlinear}, using the intensity multipoles of the CMB, one can separate the intrinsic monopole $T^\text{cmb}_0$ (the black line with a blackbody spectrum), from the  motion-induced spectral distortions (the red and blue lines) to overcome this degeneracy. 
	
	Therefore, let us now turn our attention to the intensity multiples.
	According to Eq. \eqref{monopole_intensity}, a generic  observer who is  in motion with the CMB rest frame and observing  $T^\text{eff}$ in every line of sight, will see a monopole intensity equal to 
	
	\begin{align}\label{intensity_monopole_T_eff}
	a^{I}_{00}(\nu) &=\tilde{B}_\nu(T^\text{eff}_0) a^{T^\text{eff}}_{00} + \tilde{G}_\nu (T^\text{eff}_0)\sum_{m} \frac{1}{2 \sqrt{\pi}} | a^{T^\text{eff}}_{1 m}|^2.
	\end{align}
	Substituting $T^\text{eff}_0, a^{T^\text{eff}}_{00}$ and $a^{T^\text{eff}}_{1 m}$ from Eqs. \eqref{T_eff_monopole}-\eqref{T_eff_monopole_simp} and expanding to second order in $\beta$ gives 
	\begin{align}
	a^{I}_{00}&(\nu) =\tilde{B}_\nu(T^\text{cmb}_0) a^{T^\text{cmb}}_{00} \\\nonumber
	+& \tilde{B}^{(20)}_\nu (T^\text{cmb}_0) [\beta^2 a^{T^\text{cmb}}_{00} + 4\sqrt{\pi} \beta \sum_{m}a^{T^\text{cmb}}_{1 m} Y_{1m}(\betahat)]\\\nonumber
	+&\tilde{G}_\nu (T^\text{cmb}_0)\sum_{m} \frac{1}{2 \sqrt{\pi}} | a^{T^\text{cmb}}_{1 m}|^2.
	\end{align}
	where $\tilde{B}^{(20)}_\nu(T^\text{cmb}_0)=\frac{1}{6}(g(x)-3)\tilde{F}_\nu(T^\text{cmb}_0)$. This is precisely Eq. 10 in the letter, with the addition of the nonlinear contribution of the intrinsic dipole to the observed intensity monopole. This equation can be simplified as

	\begin{align}
	I_\nu^{(0)}& =\tilde{B}_\nu(T^\text{cmb}_0)T^\text{cmb}_0 \\\nonumber
	+& \tilde{B}^{(20)}_\nu (T^\text{cmb}_0) [\beta^2 T^\text{cmb}_{0} + 2\beta d (\dhat \cdot \betahat)]\\\nonumber
	+&\tilde{G}_\nu (T^\text{cmb}_0)\frac{d^2}{3}.
	\end{align}
	Even though it might not be clear from the simplified version, it is worth mentioning that the spectral distortion $\tilde{B}^{(20)}_\nu(T^\text{cmb}_0)$ is generated because the boost is changing the temperature monopole in equation  \eqref{intensity_monopole_T_eff} as well as the dipole (compare the expression in brackets with Eq. \eqref{T_eff_monopole_simp}). 
	
	Fig. \ref{fig:monopolecorrectionsd30uknonlinear} shows the typical amplitude and frequency dependence of all these terms.
	From Fig. \ref{fig:monopolecorrectionsd30uknonlinear}, it is obvious that since the frequency dependence of the non-linear contribution of the dipole to the monopole is different from the motion-induced distortions, one could in principle use this component to measure the intrinsic dipole independently. However, the non-linear contribution of a dipole smaller than $d=300\mu$K is lower than the sensitivity of \emph{PIXIE} and therefore, it cannot be easily detected. 
	This is why the letter focuses on an alternative strategy to determine the intrinsic dipole.

	The analogous expression for dipole intensity  from Eq. \eqref{higher_multipole_intensity} reads:
	\begin{equation}
	\delta I^{(1)}_\nu(\gammahat)  = \tilde{F}_\nu(T^\text{cmb}_0)( \beta T^\text{cmb}_0 \betahat + d\dhat) \cdot \gammahat.
	\end{equation}
	Here, since we only have an intrinsic monopole and dipole, all the Wigner 3j elements in Eq. \eqref{Gaunt} (nonlinear terms) have vanished. 
	
	Now, using these general formulas, let's consider  a few different examples to illustrate the difference between the intrinsic and kinematic dipole components and their effect on the intensity monopole and dipole. 
	
	\subsubsection{Pure Kinematic Dipole}
	The current assumption in the modeling of the CMB is that the whole dipole is of a kinematic origin, and the intrinsic part is identically zero; that is $d=0$. In this case the observed temperature monopole and dipole will be equal to 
	
	\begin{equation}
	T^\text{eff}_0 = (1-\frac{\beta^2}{6}) T^\text{cmb}_0
	\end{equation}
	\begin{equation}
	\delta T^\text{eff}_{(1)}(\gammahat)  = \beta T^\text{cmb}_0 (\betahat \cdot \gammahat),
	\end{equation}
	and the observed intensity monopole and dipole will be 
	\begin{align}\label{kinematic_monopole}
	I_\nu^{(0)}& =\tilde{B}_\nu(T^\text{cmb}_0) T^\text{cmb}_{0} + \beta^2 \tilde{B}^{(20)}_\nu (T^\text{cmb}_0) T^\text{cmb}_{0} 
	\end{align}
	\begin{align}\label{kienmatic_dipole}
	\delta I_\nu^{(1)}(\gammahat)& =\tilde{F}_\nu(T^\text{cmb}_0) \beta T^\text{cmb}_{0} (\betahat \cdot \gammahat).
	\end{align}
	So, if the dipole has a kinematic origin, it will induce a spectral distortion $\tilde{B}^{(20)}_\nu (T^\text{cmb}_0)$  in the observed intensity monopole as well. 
	\subsubsection{Pure Intrinsic Dipole}
	On the other hand, if the dipole is completely intrinsic ($\beta=0$) then one would have 
	\begin{equation}
	T^\text{eff}_0 = T^\text{cmb}_0,
	\end{equation}
	\begin{equation}
	\delta T^\text{eff}_{(1)}(\gammahat)  = d (\dhat \cdot \gammahat),
	\end{equation}
	for the temperature monopole and dipole and 
	\begin{align}
	I_\nu^{(0)}& =\tilde{B}_\nu(T^\text{cmb}_0) +\tilde{G}_\nu (T^\text{cmb}_0)\frac{d^2}{3},
	\end{align}
	\begin{align}
	\delta I_\nu^{(1)}(\gammahat)& =\tilde{F}_\nu(T^\text{cmb}_0) d (\dhat \cdot \gammahat)
	\end{align}
	for intensity. Comparing these with Eqs. \eqref{kinematic_monopole} and \eqref{kienmatic_dipole}, shows that the  distortions in the intensity monopole due to an intrinsic dipole have a completely different spectral shape than the ones induced by motion. 
	
	\subsubsection{Equal Parts Intrinsic and Kinematic Dipole}
	We showed that a pure intrinsic and a pure kinematic dipole leave different signatures on the observed intensity monopole. But in reality the observed dipole will be a combination of these two cases. We show that even if the intrinsic and kinematic dipole vectors are exactly the same, it is still possible to disentangle them using the  intensity monopole and dipole. 
	
	Let's assume that the intrinsic and kinematic dipoles both point in the same direction and contribute equally to the observed dipole. For illustrative purposes, let's use the notation $\vec{\tilde{\hm \beta}} = \tilde{\beta} T^\text{cmb}_0 \hat{\tilde{\hm \beta}} $ for the amplitude and direction of the overall dipole and set $\beta = \frac{1}{2} \tilde{\beta}$ and $d = \frac{1}{2} \tilde{\beta}  T^\text{cmb}_0$ and $\dhat = \betahat = \hat{\tilde{\hm \beta}} $. In this case the observed temperature monopole and dipole will be 
	\begin{equation}
	T^\text{eff}_0 = (1-\frac{\tilde{\beta}^2}{8}) T^\text{cmb}_0,
	\end{equation}
	
	\begin{equation}
	\delta T^\text{eff}_{(1)}(\gammahat)  =  \tilde{\beta} T^\text{cmb}_0 (\hat{\tilde{\hm \beta}}  \cdot \gammahat),
	\end{equation}
	and for the intensity we have 
	\begin{align}
	I_\nu^{(0)} =&\tilde{B}_\nu(T^\text{cmb}_0)T^\text{cmb}_0 + \frac{3}{4} \tilde{\beta}^2 \tilde{B}^{(20)}_\nu (T^\text{cmb}_0)  T^\text{cmb}_{0} \\\nonumber
	&+\tilde{G}_\nu (T^\text{cmb}_0)\frac{1}{12}(\tilde{\beta}T^\text{cmb}_0)^2,
	\end{align}
	
	\begin{align}
	\delta I_\nu^{(1)}(\gammahat)& =\tilde{F}_\nu(T^\text{cmb}_0) \tilde{\beta} T^\text{cmb}_0 (\hat{\tilde{\hm \beta}} \cdot \gammahat).
	\end{align}
	Comparing this with the pure kinematic case shows that the existence of an intrinsic dipole component introduces an extra spectral distortion proportional to  $\tilde{G}_\nu (T^\text{cmb}_0)$, but more importantly it changes the amplitude of the $\tilde{B}^{(20)}_\nu (T^\text{cmb}_0)$  distortion. In other words, by measuring the amplitude of the observed overall dipole, we know how much $\tilde{B}^{(20)}_\nu (T^\text{cmb}_0)$ spectral distortion we should expect in the monopole intensity (Eq. \eqref{kinematic_monopole}), and therefore any deviations from that will be due to an intrinsic dipole component.

	These simple examples show that it is a misconception to consider that the effect of our motion with respect to a CMB rest frame is simply to produce a temperature dipole, indistinguishable from any intrinsic dipole the CMB may possess.

\end{document}